\documentclass[
    ,final            
  ]
  {aipproc}

\layoutstyle{6x9}

\begin{document}

\title{Lower Redshift Analogues of the Sources of Reionization}


\author{Michael Rauch}{
  address={Carnegie Observatories, 813 Santa Barbara Street, Pasadena CA 91101, USA}
}



\begin{abstract}
Known populations of  QSOs appear to fall short of producing the ionizing flux 
required for re-ionizing the universe. The alternative, galaxies as sources of ionizing photons, suffers from the problem
that known types of galaxies are almost completely opaque to ionizing photons.  For reionization to happen, either large numbers
of (largely undiscovered) sources are required, or the known populations of galaxies need to have had a much larger escape fraction for ionizing radiation in the past. 

We discuss recent discoveries of faint $z\sim 3$ Lyman $\alpha$ emitters with
asymmetric, extended Ly$\alpha$ emission regions, which apparently are related to interacting galaxies.
The unusually shaped Ly$\alpha$ line profiles and the underlying stellar populations of these objects suggest
the presence of damaged gaseous halos, infall of gas, tidal or stripped stellar features and
young populations of hot stars
that would all be conducive to the release of ionizing radiation into intergalactic
space. As galaxy interactions and mergers increase with redshift, these effects can only become more
important at earlier times, and so these interacting $z\sim3$
objects may be late, lower redshift analogues of the sources of reionization.
\end{abstract}

\maketitle


\section{Reionization by Galaxies}

The known population of QSOs, thought to be the main source of ionizing photons throughout most of the history of the universe, falls short of maintaining
the ionization of the intergalactic medium at redshifts beyond $z\sim 4$ (e.g., \cite{rau97}), whereas the universe is
known to be fully ionized back to at least $z\sim 6$ (e.g., \cite{bec07}). Ionizing flux escaping from galaxies is often
invoked to make up for the missing photons. The highest redshift galaxy surveys are  finding increasing numbers of $z>7$ galaxies. However, to reach the number density of photons necessary for re-ionization may require either the existence of
large numbers of (still) invisible sources (e.g., Trenti, this conference), or rather large escape fractions for those types of galaxies currently observable (e.g., Finkelstein, this conference).
The challenge posed by these conditions  is exacerbated by a paucity of credible observations of any galaxies emitting a significant amount of ionizing radiation (see, e.g., \cite{van10} for a recent discussion). This may suggest that yet other types of sources may be responsible for
reionization ("hidden" AGN, mini-halos), or that the galaxies at those lower redshifts currently accessible to observation
for some reason are no longer  leaking ionizing photons.


With ordinary galaxies failing to ionize the universe, it is instructive  to briefly consider possible conditions that
may facilitate the release
of ionizing radiation. These conditions generally fall into the following three categories:
\begin{enumerate}
\item enhanced production of ionizing photons, e.g., by hotter, perhaps metal poor stars, or 
production of harder photons that experience a lower optical depth.

\item creation of holes in the otherwise optically thick galactic disks or halos, through which ionizing radiation
may escape, possibly through interactions that damage a gaseous halo, galactic winds, or through ionization by an AGN.

\item the formation in, or transport of the stellar sources of ionizing radiation into less optically
thick environments. Examples may be any kind of intra-halo star formation, e.g., in globular clusters or tidal tails. 

\end{enumerate}

In the absence of the difficult, direct detections of leaking ionizing continuum, the prime observational diagnostic of the above features would be  the Ly$\alpha$ emission from neutral hydrogen.
A Ly$\alpha$ emission line from a gaseous halo that is leaking ionizing radiation may be spatially asymmetric, and possibly offset from the underlying sources of the ionizing 
continuum.
Density and velocity disturbances of the halo gas, caused by galaxy interactions, could produce such asymmetries, as could multiple
sources of ionizing radiation moving about in a joint gaseous halo. The presence of the latter may also be evident from broad
band detections of multiple stellar continuum objects, perhaps in the form of disturbed stellar populations, e.g., tidal features, and young stellar populations, the formation of which may have been triggered by interactions.

\section{Detecting candidate sources of ionizing photons from searches for faint Ly$\alpha$ emission}

Probable detections of some of the above phenomena have recently come from ultra-deep spectroscopic blind searches for Ly$\alpha$ emission at redshift $\sim 3$.
In a reversal of the usual search strategy for Ly$\alpha$ emission (narrow band imaging combined with spectroscopic followup),
a spectroscopic blind survey detects Ly$\alpha$ emission line features in a randomly positioned, deep 2-dimensional long slit spectrum. Underlying continuum sources can be inspected in existing space-based imaging at the slit position.
Because of its superior ability to suppress the sky background, this method is ideally suited  for finding faint galaxies (and  has been able to uncover a large number density of $z\sim3$ Ly$\alpha$ emitters with luminosities down to $5\times10^{-3}\  L_*$; \cite{rau08}). As a by-product, this approach also uncovers very extended, low surface brightness Ly$\alpha$
emitters that would have escaped traditional searches.

In a deep 61.4 h long slit spectrum, obtained with the Magellan LDSS3 instrument in the HUDF, we have discovered three such extended peculiar Ly$\alpha$ emitters in a volume of 2056 h$_{70}^{-3}$ Mpc$^{-3}$ (comoving). The 2-d spectrum shows
Ly$\alpha$ emission by these objects across several tens of kpc, with total fluxes on the order of a few $\times 10^{-17}$ erg cm$^{-2}$s$^{-1}$. While one
of the objects turned out to be an obscured QSO (and will be described in a future publication), the other two which we shall describe here appear to be related to interacting galaxies, with no obvious signs of the presence of an AGN.

\begin{figure}
  \includegraphics[height=.35\textheight]{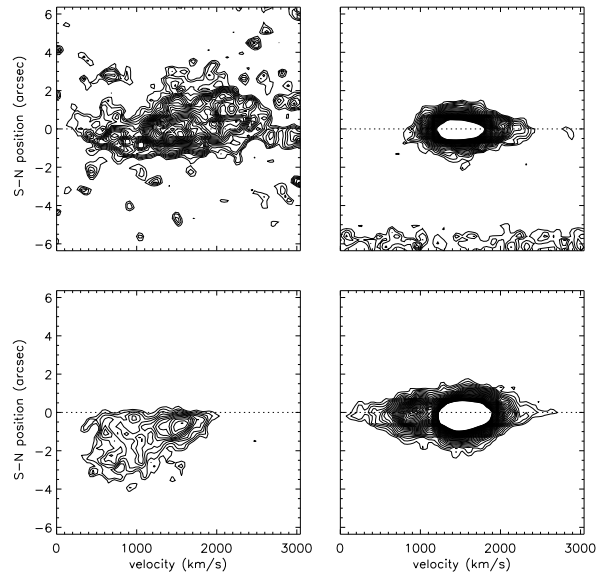}
  \caption{A comparison between the appearance of the Ly$\alpha$ emission line regions of the two extended, asymmetric objects
(left column) discussed here, and two more normal, compact Ly$\alpha$ emitters from galaxies that could have been detected by either  Lyman break or narrow band filter surveys (right column). The latter, much  more typical galaxies appear to have spatially symmetric halos, a more orderly velocity field, and can be explained by single sources of ionization.
The galaxy in the bottom left panel is referred to below as "case I", the one in the top left panel as "case II".\label{compares}} 
\end{figure}

To illustrate the differences between these extended, asymmetric emitters and standard  Lyman $\alpha$ emitting galaxies we
show in fig. \ref{compares} parts of the spectra of the two extended emitters discussed here in the panels on the left, and two other bright, previously
known Lyman break galaxies with Ly$\alpha$ emission on the right side. The differences are striking, in that the Lyman break
galaxies show very compact, symmetric, spatial Ly$\alpha$ profiles,  and spectral profiles with the usual red-dominated 
peak. Models of Ly$\alpha$ radiative transfer (\cite{dij06}, \cite{bar09}) show that this pattern is
consistent with a single compact source of ionizing photons embedded in an HI halo highly optically thick to ionizing and Ly$\alpha$ photons. 
In contrast, the extended emitters are spatially asymmetric, with a complex, multi-component spectral profile.
Evidence to be described below suggests that the latter objects  are  gaseous halos where the velocity and density structure of the gas has been disturbed by interactions and multiple internal sources of ionizing and Ly$\alpha$ photons.

\subsection{Case I: radiation escaping from an interacting galaxy at z=3.34}

\begin{figure}
  \includegraphics[height=.22\textheight]{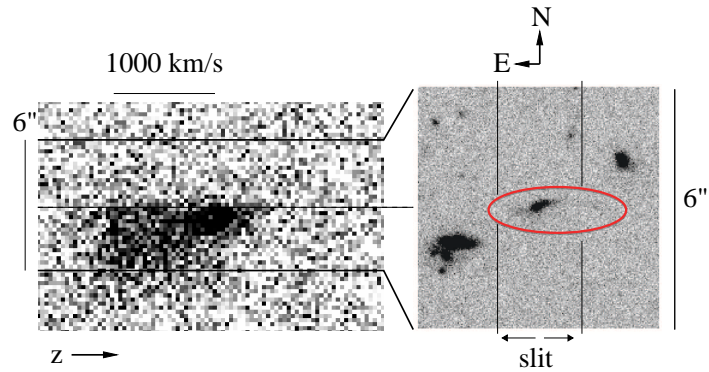}
  \caption{2-d LDSS3 spectrum of the Ly$\alpha$ emission line region of the 3.34 emitter (left) and a V band image from
the GOODS-S survey (\cite{gia04}) of the underlying galaxy. The emitter corresponds to a V=27 galaxy with stellar tails sticking out horizontally.
\label{sys1}}
\end{figure}

\begin{figure}
  \includegraphics[height=.18\textheight]{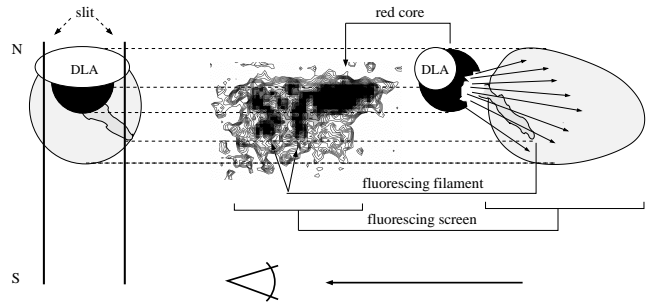}
  \caption{Diagram explaining how the spectral pattern of Ly$\alpha$ emission in the z=3.34 object relates to the
escape of ionizing radiation: frontal view through the slit (left), and view perpendicular to the line-of-sight (center-right). The compact Ly$\alpha$ emission line propagating from the center of the galaxy ("red core") is partly absorbed by an optically thick damped Ly$\alpha$ region. The
fluorescent screen and filament represent external gaseous structures falling into the galaxy on the far side from the observer. They are irradiated by ionizing photons escaping from a hole in the back of the gaseous halo and fluoresce in Ly$\alpha$. 
\label{sys1_scheme}}
\end{figure}

The first case shows a "fan" of low surface brightness emission extending over about 35 kpc (proper; fig. \ref{sys1}). The underlying
galaxy is a disturbed V=27 object with apparent stellar tails sticking out on two sides. The relatively
complex spectral and spatial shape of the Ly$\alpha$ emission can be interpreted as being caused by gas falling in from the back onto the galaxy (fig. \ref{sys1_scheme}).
The in-falling nature of the gas can be ascertained from the fact that the blue-shifted emission is sharply cut out on one side by the foreground damped Ly$\alpha$ absorption from the galaxy. The simple double humped structure of the emission line 
may arise when ionizing radiation directly impacts a filament of HI gas, "reflecting" part of the
energy back in the form of Ly$\alpha$, as opposed to the more highly modified line profiles expected for Ly$\alpha$ emerging
through multiple scattering from the depth
of an intact halo.  
The ionized photons produced by the stars in that galaxy can account for the entire observed flux in Ly$\alpha$ if we
assume  that about 50\% of the photons escape the halo to hit the external HI. 
In the absence of other possible sources of ionization this may be an actual observation of a massive escape of stellar ionizing radiation
from an interacting galaxy. The presence of an AGN, as an alternative source of ionizing photons and reduced opacity, cannot be ruled out, but we have no positive evidence for it.
In the absence of an AGN, the ionizing photons will have to escape through actual holes in the HI cocoon of the galaxy,
or perhaps from a partly stripped, tidal tail surrounded by regions of lower Lyman limit opacity.
The object is also remarkable in that part of the fluorescent Ly$\alpha$ emerges from a thin filamentary structure of in-falling gas connected to the galaxy. The filament may be a plausible detection of a cold accretion inflow, as it does not seem to line up with any stellar features (\cite{rau11}).

\subsection{Case II: filamentary structure and young, in situ star formation in a Milky-Way sized halo of interacting galaxies at z=2.63}

\begin{figure}
  \includegraphics[height=.18\textheight]{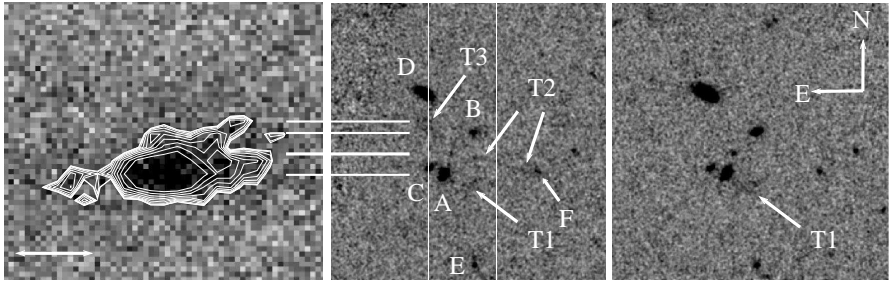}
  \caption{The z=2.63 Ly$\alpha$ emitter. 2-d spectrum (LHS panel), HST ACS B-band (central panel),
and V-band image (RHS panel). The image size is $10"\times10"$. The spectrum shows a filamentary patch of Ly$\alpha$ emission. In the broad band images, several blueish galaxies
are visible, together with patchy or filamentary  emission features. Of these, T1 appears to be a continuum 
feature, whereas T2 and T3 may be dominated by Ly$\alpha$ emission. T2 could be a tidal tail or a turbulent stripped wake behind a galaxy moving through the gaseous halo. Recent star formation may explain the apparently large equivalent width (\cite{rau12}).\label{2ndsys}}
\end{figure}

The second object is at z=2.63, and the spectrum again shows a complex and extended, partly filamentary emission line (\cite{rau12}; left-most panel in fig. \ref{2ndsys}). A second filament ("T2" central panel; mostly outside
of the spectrograph slit) appears
in the B-band image  but not in the V-band, suggesting that its flux may also be dominated by Ly$\alpha$ line radiation, however, with a very large equivalent width. 
The underlying continuum sources
here  again exhibit several diffuse emission features that may be of tidal origin. Most galaxy redshifts cannot be established with absolute certainty because of the faintness of the objects, and the multiple published photometric redshifts for several objects are mutually incompatible. We propose that this may be due to
the presence of a very young (a few $10^6 $yr) stellar population in some of the objects (\cite{rau12}).
In the current case we cannot discern direct evidence for escaping ionizing radiation, but the filamentary features suggest that
the galaxies have interacted in ways that produced an extended stellar distribution, with possible tidal tails or stripped wakes
exhibiting signs of recent star formation. The large Ly$\alpha$ equivalent widths inferred and the fact that the stars appear
to be forming in situ in a metal-poor gaseous halo suggests that these may be hot  stars
with enhanced yields of ionizing photons (e.g., \cite{rai10}, \cite{ino11}). The location of the stars
outside of the optically thickest regions of the individual galaxies would also favor the escape of ionizing radiation.
Thus, this object meets several of the hypothetical criteria outlined above for the galactic sources of ionizing photons.

\section{Conclusions}

We have discussed some astrophysical conditions that could turn galaxies into sources of ionizing radiation during the epoch
of reionization, and we described two recently discovered Ly$\alpha$ emitting galaxies that satisfy several of these conditions,
including an apparently leaky gaseous halo, and possible extragalactic, intra-halo star-formation from a metal poor reservoir of gas (apparently in the form of tidal tails or turbulent wakes). The connection of these features to galaxy interactions
suggest that, as the merger rate increases with redshift, such objects
will be much more common at the epoch of reionization. The extended $z\sim 3$ emitters discussed here may just be
the lower redshift remnants of this population. In contrast, the symmetric, compact Ly$\alpha$ emitters or Lyman break galaxies represent the
bulk of the $z\sim 3$ star-forming galaxies.  In the absence of recent interactions, these objects, with their more spatially concentrated star-formation, more mature stellar populations, and their ionizing radiation from a dominant central source being trapped by an inviolate gaseous halo,  may be less likely to contribute to the ionization of the universe.

We conclude that interactions between galaxies  may be the missing ingredient that turns high redshift galaxies into the anticipated sources of ionizing radiation.

In neither one of the two peculiar Ly$\alpha$ emitter cases can we exclude the presence of an AGN, which could provide alternative explanations for both the high Ly$\alpha$  
widths observed, and the escape of ionizing photons.
However, taking into account that there is already another QSO in the same (small) survey volume, the assumption, that all three objects could reflect photonionization by internal QSOs,  would raise the number density for QSOs at $z\sim3$ to $\sim 1.5\times10^{-3}h_{70}^3$Mpc$^{-3}$. Such a high value in itself may have interesting consequences for the budget of ionizing photons.



\begin{theacknowledgments}
I would like to thank 
my collaborators George Becker, Martin Haehnelt, Jean-Rene Gauthier, Swara Ravindranath, and Wal Sargent, and acknowledge useful discussions with Sebastiano Cantalupo, Bob Carswell, Hsiao-Wen Chen, Jeff Cooke,  
Li-Zhi Fang, Pat McCarthy, Masami Ouchi, and Francois
Schweizer. I further thank the staff of the Las Campanas and the Keck Observatories for their help with the
observations, the IoA in Cambridge and the Raymond and Beverley Sackler Distinguished Visitor program for hospitality and support, and the NSF
for funding through grant AST-1108815.
\end{theacknowledgments}



\bibliographystyle{aipproc}   



\end{document}